# Suppression of Superconductivity and Nematic Order in $Fe_{1-y}Se_{1-x}S_x$ (0 ≤ x ≤ 1, y ≤ 0.1) Crystals by the Anion Height Disorder

Aifeng Wang,[†,&,*] Ana Milosavljevic,[$] AM Milinda Abeykoon,[§] Valentin Ivanovski,[‡] Qianheng Du,[†,l,#] Andreas Baum,[¶,/] Eli Stavitski,[§] Yu Liu,[†,▽] Nenad Lazarevic,[$] Klaus Attenkofer[§,⊥] Rudi Hackl,[¶,/] Zoran Popovic[$,‖] and Cedomir Petrovic[†,l,*]

[†]Condensed Matter Physics and Materials Science Department, Brookhaven National Laboratory, Upton 11973 New York USA

[$]Center for Solid State Physics and New Materials, Institute of Physics Belgrade, University of Belgrade, Pregrevica 118, 11080 Belgrade, Serbia

[§]National Synchrotron Light Source II, Brookhaven National Laboratory, Upton, New York 11973, USA

[‡]Vinca Institute of Nuclear Sciences, University of Belgrade, Belgrade 11001, Serbia

[l]Materials Science and Chemical Engineering Department, Stony Brook University, Stony Brook 11790 New York USA

[¶]Walther Meissner Institut, Bayerische Akademie der Wissenschaften, 85748 Garching, Germany

[/]Fakultät für Physik E23, Technische Universität München, 85748 Garching, Germany

[‖]Serbian Academy of Sciences and Arts, Kneza Mihaila 35, Belgrade 11000, Serbia

**ABSTRACT:** Connections between crystal chemistry and critical temperature $T_c$ have been in the focus of superconductivity, one of the most widely studied phenomena in physics, chemistry and materials science alike. In most Fe-based superconductors materials chemistry and physics conspire so that $T_c$ correlates with the average anion height above the Fe plane, i.e. with the geometry of the $FeAs_4$ or $FeCh_4$ (Ch = Te, Se, S) tetrahedron. By synthesizing $Fe_{1-y}Se_{1-x}S_x$ (0 ≤ $x$ ≤ 1, $y$ ≤ 0.1) we find that in alloyed crystals $T_c$ is not correlated with the anion height as most other Fe superconductors. Instead, changes in $T_c(x)$ and tetragonal-to-orthorhombic (nematic) transition $T_s(x)$ on cooling are correlated with disorder in Fe vibrations in direction orthogonal to Fe planes, along the crystallographic $c$-axis. The disorder stems from the random nature of S substitution, causing deformed $Fe(Se,S)_4$ tetrahedra with different Fe-Se and Fe-S bond distances. Our results provide evidence of $T_c$ and $T_s$ suppression by disorder in anion height. The connection to local crystal chemistry may be exploited in computational prediction of new superconducting materials with FeSe/S building blocks.

## INTRODUCTION

The question if some specific crystal structure and bonding situation can facilitate superconducting pairing intrigued chemical research ever since the early years after discovery of superconductivity.[1-10] Hence, the connection between superconducting critical temperature ($T_c$) with aspects of crystal structure is crucial, yet poorly understood in all superconductors. This applies in particular for Fe- and Cu-based high-$T_c$ superconductors; the former also feature electronic nematic coupled with structural orthorhombic transition ($T_s$) above $T_c$.[6-16] Fe superconductor materials crystallize in different space groups but share a common local structure feature: tetrahedrally coordinated Fe atoms.[6-12] One important empirical discovery for the future materials design is that in most Fe-based superconductors the maximal $T_c$ correlates with the average anion height above the Fe plane; the height depends on the geometry of the $FeAs_4$ or $FeCh_4$ (Ch = Te, Se, S) tetrahedron.[7] The geometry also regulates the correlation strength due to average Fe-As(Ch) hybridization, pointing to a spin fluctuation mechanism of pairing that governs the magnitude of $T_c$.[7,12]

Interestingly, critical temperatures in FeSe conform to anion height curve only at high pressure when $T_s$ is fully suppressed.[7] A dome-like magnetic phase supersedes the nematic order in the absence of chemical disorder in pressurized FeSe and $T_c$ = 37 K is obtained at 6 GPa; in contrast no magnetic order was found and only weak changes in the $T_c$ were detected when FeSe is subjected to combined perturbation of chemical pressure and disorder for sulfur substitution of up to 20 % on Se atomic site.[17-19] An abrupt change of the superconducting gap structure points to two distinct pairing states as S substitutes Se across the nematic critical point (NCP) around $x$ = 0.17.[20-22] Sulfur substitution in FeSe is expected to simply suppress electronic correlations associated with Fe $d_{xy}$ orbitals.[23]

Consequently, there should be a smooth change of $T_c$ between FeSe (~ 10 K) to FeS (~ 5 K).[12]

However, due to the complex Fe-Se/S composition-temperature phase diagram and the unstable tetragonal phase which is difficult to synthesize, the clear connection of $T_c$, $T_s$ with $x$ and with crystal structure-related parameters in FeSe$_{1-x}$S$_x$ (0 ≤ $x$ ≤ 1) is still unknown.[24-26] In contrast to the complex copper oxides or other Fe-based superconductors, simpler crystal chemistry of binary iron chalcogenides might allow for deeper insight. To address this challenge, we present the entire progression of crystallographic changes as superconducting $T_c$ evolves from FeSe to FeS. With mixed S/Se occupancy of the anion atomic site, we observe Fe vacancy defects of up to about 10 %. We find that $T_c(x)$ is unrelated to the average anion height for large region of $x$ where $T_c$ is found to be connected with the variation of the anion height induced by the fixed Fe-Se and Fe-S bonds and thereby induced disorder in Fe vibrations along the crystallographic *c*-axis, orthogonal to the Fe plane.

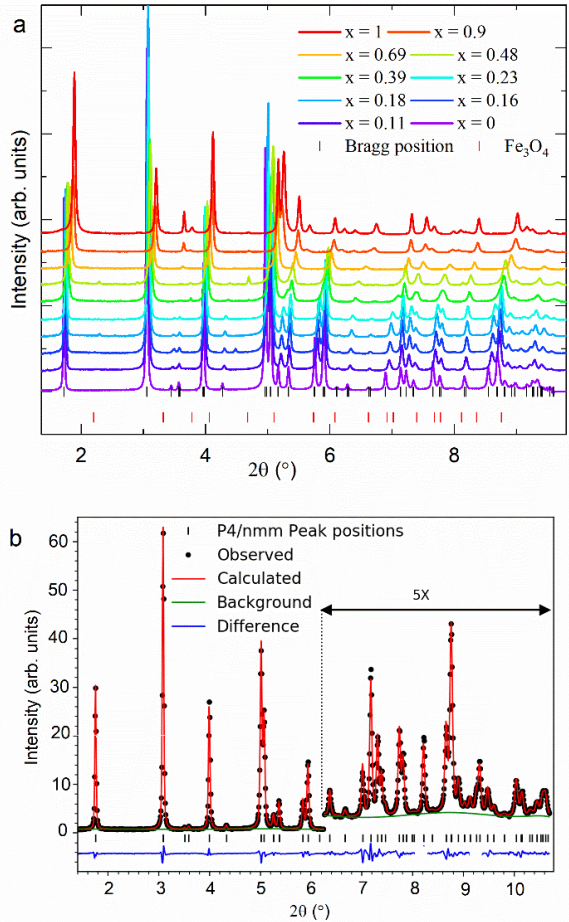

**Figure 1**. Average crystal structure from synchrotron powder XRD patterns. (a) Powder XRD patterns of Fe$_{1-y}$Se$_{1-x}$S$_x$ for (0 ≤ $x$ ≤ 1, $y$ ≤ 0.1). The top vertical tick marks represent FeSe Bragg reflections in the *P4/nmm* space group whereas bottom show reflections of Fe$_3$O$_4$. The small Fe$_3$O$_4$ peaks visible for some $x$ are due to oxidation during pulverization of air-sensitive crystals for 1 mm polyamide capillary loading. (b) Example of the Rietveld refinement of the background subtracted data for Fe$_{0.93}$Se$_{0.89}$S$_{0.11}$.

Our results show local deformations of FeCh$_4$ tetrahedra for most $x$ in Fe$_{1-y}$Se$_{1-x}$S$_x$ (0 ≤ $x$ ≤ 1) crystals and provide evidence of $T_c$ suppression related to disorder of Fe vibrations along the crystallographic *c*-axis. Since substantially higher superconducting $T_c$'s can be induced in FeSe by intercalating molecular spacer layer or by nanofabrication of ultrathin FeSe layers on SrTiO$_3$ substrates,[27-29] evidence of $T_c$ suppression related to disorder in particular vibrations of atoms in the unit cell could be used to chemically tailor not only critical temperature in intercalated FeSe crystals or few-layer samples but also in novel superconductors with tetrahedral Fe-Ch or Fe-As building blocks in the unit cell.[10]

EXPERIMENTAL SECTION

Single crystals of Fe$_{1-y}$Se$_{1-x}$S$_x$ with 0 ≤ $x$ ≤ 0.24 were grown using an eutectic mixture of the KCl and AlCl$_3$ as the transport agent.[30] Namely, Fe, Se, S, KCl, and AlCl$_3$ powders were mixed together in the ratio Fe: Se: S: KCl: AlCl$_3$ = 1.1: 1-x: x: 1: 2. After grinding in a mortar, the mixture was sealed in a quartz tube of 12 cm length and 1.4 cm inner diameter. Single crystals were grown in a horizontal tube furnace with a temperature gradient of 150 °C. Samples were very slowly heated to 420 °C in two weeks and kept for 2 months in the gradient. Shiny crystals with typical dimensions of 3 x 2 x 0.3 mm$^3$ were picked up on the cool end. Single crystals of Fe$_{1-y}$Se$_{1-x}$S$_x$ with 0.39 ≤ $x$ ≤ 1 were synthesized by deintercalation of potassium from the corresponding K$_{0.8}$Fe$_{2-y}$(Se$_{1-x}$S$_x$)$_2$ single crystals using the hydrothermal reaction method.[31,32] K$_{0.8}$Fe$_{2-y}$(Se$_{1-x}$S$_x$)$_2$ crystals were mixed with Fe powder, selenourea powder and sulfourea pieces and weighted in Fe: Se: S = 1.1: 1-x: x ratio, where Se: S atomic ratio was identical to crystals. Then, 5mmol of the mixture, several pieces of K$_{0.8}$Fe$_{2-y}$(Se$_{1-x}$S$_x$)$_2$ single crystals, 2 mmol Fe pieces, 0.1g NaOH, and 5ml deionized water were loaded to a 25 ml stainless steel autoclaves with Teflon liner. Additional Fe pieces were added to keep the reducing atmosphere. The autoclave was tightly sealed and heated to 140 °C in 3h and kept at that temperature for 72 h in a small box furnace. Single crystals can be obtained by rinsing the product using deionized water and alcohol and after overnight drying in pumped vessel.

The element analysis was performed using an energy- dispersive X-ray spectroscopy (EDX) in a JEOL LSM-6500 and in JEOL 7600F scanning electron microscopes with about 2% accuracy.

Synchrotron powder XRD was measured on pulverized crystals using 74.69 keV (0.166 Å) synchrotron radiation of National Synchrotron Light Source 2 (NSLS 2) at beamline 28-ID-1. Sample to detector distance was 1216.272 mm. The Rietveld analysis was carried out using the GSAS-II software package.[33] The X-ray absorption near-edge structure (XANES) and extended X-ray absorption fine- structure (EXAFS) experiments were obtained by mixing pulverized single crystals and BN powder uniformly, and then pressing the mixture into pellets. The experiments were carried out at room temperature in the 8-ID beam line of the NSLS 2 at Brookhaven National Laboratory (BNL) in the transmission mode. Data were processed and analyzed using the ATHENA and ARTEMIS software programs.[34] The AUTOBK code was used to normalize the absorption coefficient and to separate the EXAFS signal, $\chi(k)$ from the atom-absorption background. The extracted EXAFS signal, $\chi(k)$, was weighed by $k^2$ to emphasize the high-energy oscillation and then Fourier-transformed in a $k$ range from 2 to 5 Å$^{-1}$ to analyze the data in the R space.

The investigation of the Mössbauer effect in the powdered samples, the tetragonal Fe$_{0.9}$Se$_{0.82}$S$_{0.18}$ and the tetragonal FeSe$_{0.31}$Se$_{0.69}$, was performed in transmission geometry using a

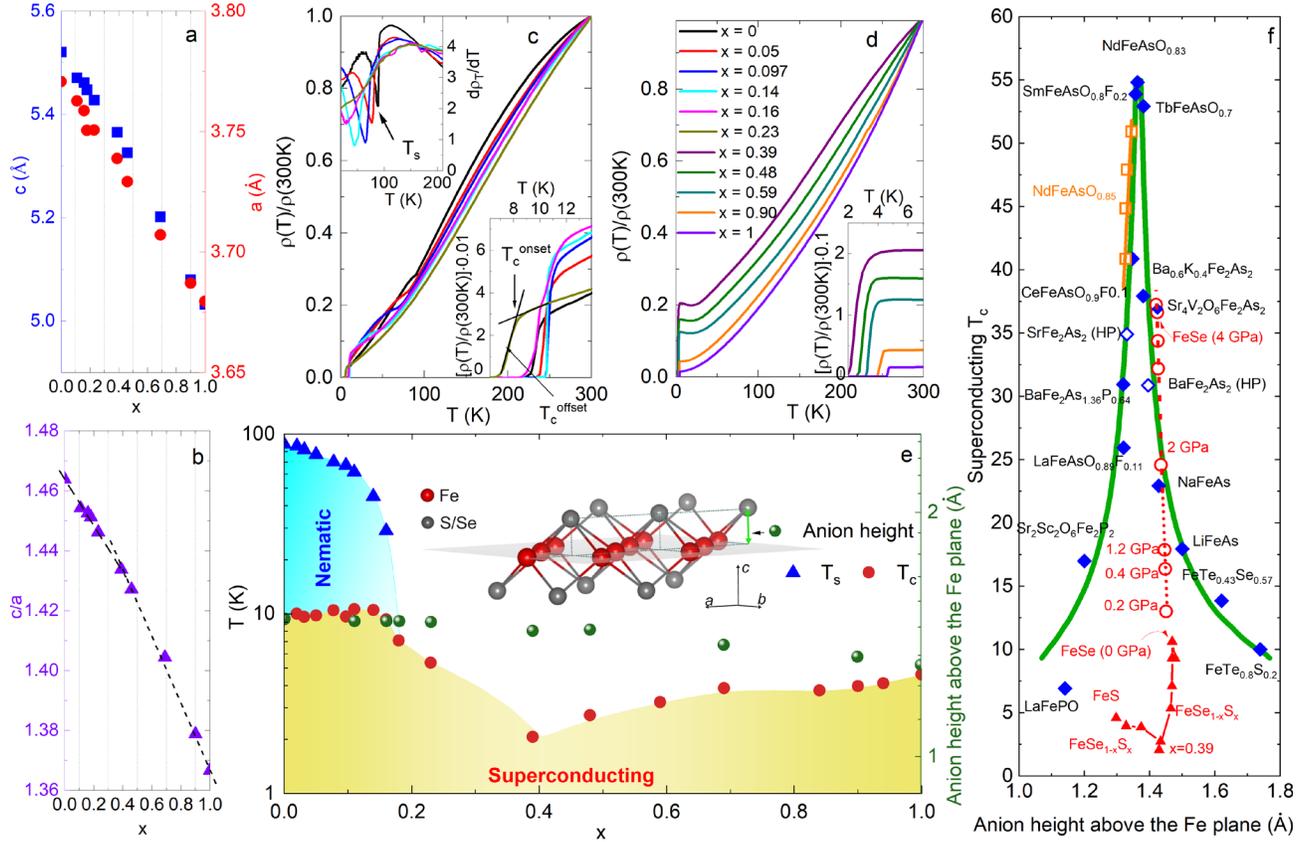

**Figure 2**. Changes of the average anion height above the iron plane in $Fe_{1-y}Se_{1-x}S_x$ ($0 \leq x \leq 1$, $y \leq 0.1$). (a,b) Lattice parameters of the crystal structure $a$, $c$ and $c/a$ ratio obtained from Rietveld refinement. Electrical resistivity for $Fe_{1-y}Se_{1-x}S_x$ (c,d) normalized to values at 300 K $\rho_T = \rho(T)/\rho(300K)$. The left inset in (c) shows the $d\rho_T/dT$ around the structural transition; structural transition temperature $T_s$ is inferred by the dip and indicated by the arrow. Legends for Fig. 2(c,d) are identical and are shown in Fig. 2d. The composition-temperature ($x$ – $T$) phase diagram for single crystals (e) with values of $T_c$, $T_s$ and the average anion height above the Fe plane, depicted in inset. Superconducting $T_c$ in iron-based superconductors (f) from Ref. 5 with $T_c$'s from this work (red triangle symbols).

[57]Co(Rh) source at the room temperature. The spectrometer was calibrated by the spectrum of a natural iron foil. The spectra were collected from samples in low and in high velocity range. All Mössbauer spectra have been examined by the Win-Normos-Site software package based on the least square method.[35] The measured isomer shift values ($\delta$) are given relative to metallic alpha iron ($\delta = 0$).

Electrical transport was measured in a Quantum Design PPMS-9 using a standard four-probe configuration with current flowing in the tetragonal plane. Magnetization was measured using a Quantum Design MPMS-XL5. Sample dimensions were measured with an optical microscope Nikon SMZ-800 with 10 $\mu$m resolution. The dimensions were measured several times, then averaged. The standard deviation was less than 2%.

Raman experiments were performed using an excitation source solid state laser emitting at 532 nm. In our scattering configuration, the plane of incidence is the $ab$ plane, with incident (scattered) light propagation along the $c$-axes. Right before being placed in vacuum samples were cleaved in the air. All measurements were performed in high vacuum ($10^{-6}$ mbar) using a KONTI CryoVac continuous Helium flow cryostat with 0.5 mm thick window. The laser beam was focused using microscope objective with x50 magnification. All spectra were corrected for Bose factor. Experiments were performed using different crystals in two different institutions: Walther Meisner Institute (WMI) and Insititute of Physics Belgrade (IPB).

RESULTS

Synchrotron X-ray powder diffraction patterns (Fig. 1a) confirm the phase purity of $Fe_{1-y}Se_{1-x}S_x$ for all investigated crystals in the range $0 \leq x \leq 1$. Typical Rietveld refinement of the average unit cell is shown in Fig. 1b on the example of $Fe_{0.93}Se_{0.89}S_{0.11}$. Details of refinement are given in Supporting information.[36] The evolution of the unit cell parameters with S substitution on Se atomic site, inferred from the Rietveld refinement of the average crystal structure (Fig. 2a) shows that both the $a$ and $c$ lattice parameters decrease monotonically as S is increased. However $a$ and $c$ evolve at a different rate at low S substitution level, in contrast to same rate at the higher level. This change manifests itself as a clear kink in the $c/a$ ratio (Fig.2b) near $x = 0.25(5)$. The simultaneous decrease of the $c/a$ ratio and the suppression of the resistivity hump with increasing $x$ (Figs. 2c,d) suggest that the average crystal structure becomes more three-dimensional as $x$ is increased. The reduction of the $c/a$ ratio is stronger in the S-substituted crystals than in pure FeSe under high pressure where the crystal structure of the superconducting phase having maximal $T_c$ at 37 K changes, and magnetic order is observed up to 40 GPa.[17,18]

The resistivity anomaly associated with the structural transition at $T_s$ in FeSe is suppressed to lower temperature by S (Fig.

2c,d) approaching T → 0 K above $x$ = 0.18.[20] The $T_c$ increases slightly with S substitution from 9.3 K in FeSe to 10.9 K around $x$ = 0.1, in agreement with previous observation,[37] whereas the transition width becomes broader and $T_c$ decreases quickly with further S substitution, reaching a minimum value of 2.1 K at $x$ = 0.4 (Fig. 2c,d). With even higher sulfur substitution $T_c$ increases slowly up to 4.6 K for $x$ = 1 (Fig. 2d). The $T_c$ values are consistent with the ones inferred from magnetic susceptibility whereas values of ρ(300 K) are shown in Table S1 in Supplementary Information.[36]

In Fe-based superconductors the highest $T_c$ values are observed for regular Fe-Ch tetrahedrons where all bond angles approach 109.5°, empirically scaling with the average Ch(As) height above the Fe plane.[7] We plot the average anion height from Rietveld refinement in Fig. 2e, along with $T_c$ and $T_s$ evolution with $x$. The $T_c$'s for FeSe and FeS in the absence of chemical substitutions do scale with change in the average anion height in the phase diagram. However, $T_c$'s for FeSe, FeS and for any of the Se/S single crystal alloys are not on the curve that connects anion height above the Fe plane with critical temperature in Fe-based superconductors; one should expect $T_c$ of about 20 K for both compounds since heights of Se and S atoms from the Fe atomic plane are 1.477(1) Å and 1.297(1) Å for FeSe and FeS, respectively. On the other hand, due to smaller atomic radius of sulfur atom, it is expected that S substitution in FeSe will bring chemical pressure effect. Next, we will discuss differences in the effects of pressure [$T_c(P)$], S substitution [$T_c(x)$] and small amount of disorder introduced by variations of KCl/AlCl$_3$ synthesis conditions[38] on $T_c$.

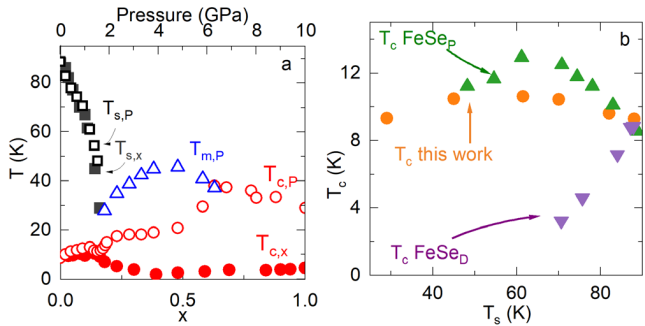

**Figure 3.** Comparison of pressure and sulfur substitution. (a) Comparison of the ground-state changes as a function of chemical and hydrostatic pressure with $x$. Filled symbols represent Fe$_{1-y}$Se$_{1-x}$S$_x$ whereas open symbols represent the temperature-pressure phase diagram of FeSe single crystals reported by Sun et al.[18] $T_c(T_p)$ in Fe$_{1-y}$Se$_{1-x}$S$_x$, pressurized FeSe (FeSe$_P$ and in FeSe disordered by different synthesis conditions (FeSe$_D$) (References.[18,38]

The evolution of the lattice parameters (Fig. 2a) is different from the observed changes of the crystal structure of FeSe under applied pressure $P$. At high pressures $2c/(a+b)$ decreases smoothly up to 4 GPa and then increases up to 9 GPa where the tetragonal crystal structure transforms to hexagonal FeSe.[39] When the applied and chemical pressure in the FeSe phase diagram are scaled by the structural transition temperature (10 GPa for full S occupancy), $T_c(x)$ and $T_c(P)$ agree reasonably well with each other (Fig. 3a) up to about the NCP.[18] Beyond the NCP there is a local minimum of $T_c(P)$ at around 1.5 GPa and magnetic order appears with the increase in $T_c$ up to 30 K whereas $T_c$ in Fe$_{1-y}$Se$_{1-x}$S$_x$ decreases to a minimum at $x$ = 0.4 and increases with further S substitution. Thus, S provides mainly chemical pressure effects for $x$ values below the NCP. This is also evident from the size of $a$ lattice parameter (Fig. 2a) that decreases by $x$ = 0.16 to about 3.75 Å, consistent with pressurized FeSe.[17] It is instructive to stay focused on the nematic region and to compare $T_c(T_s)$ induced by S substitution, pressure and disorder (Fig. 3b).[18,38,40] The $T_c(T_s)$ values in Fe$_{1-y}$Se$_{1-x}$S$_x$ and pressurized FeSe have similar dome-like behavior with the highest $T_c$ at $T_s$ ~ 60 K; in contrast $T_c$ decreases quickly with the suppression of $T_s$ in FeSe disordered by small variations in the synthesis conditions. We also note that, even though intrinsic long-range magnetic order is absent in both FeS and FeSe and superconductivity is robust to small impurity phases,[30,41-44] there are strong magnetic fluctuations and magnetic order induced by pressure.[44-47] The absence of long-range magnetic order therefore can be explained by the collapse of the effect of chemical pressure just before the critical pressure for magnetic order is reached, as evidenced by the sharp deviation from the initial slope of the $c/a$ ratio near the NCP (Fig. 2b).

Beyond average crystal structure, it is also instructive to address atomic element-specific bond distances. To shed light on the chemical bonding and the key parameters of the local crystallography of FeCh$_4$ (Ch = Se, S) tetrahedra we studied the X-ray absorption near-edge structure (XANES) and the extended X-ray absorption fine-structure (EXAFS) which provide important information on the hybridization between the Fe 3$d$ and Ch 4$p$ orbitals as well as accurate Fe-Ch bond distances.[48,49]

Typical features of the Fe K-edge XANES are denoted as $A$, $B$, $C$, whereas those of the Se K-edge are denoted as $D$, $E$ (Fig. 4a,b). The pre-peak $A$ is due to the direct 1$s$ → 3$d$ quadrupole transition to unoccupied states, with a contribution of dipole transition from the Fe 1$s$ to unoccupied Fe 3$d$ - Se 4$p$ hybrid bands.[50] The edge feature $B$ is determined by the 1$s$ → 4$p$ transitions whereas feature $C$ results from the 1$s$ → 4$p$ state with a significant admixture to the Ch $d$ states. The intensity of $A$ increases with S concentration, indicating an increase of the Fe 3$d$ - Se 4$p$ hybridization. The decrease of the intensity of structure $C$ indicates the reduction of the hybridization of Fe 4$p$ and Se $d$ states due to low Se concentration. The results for the Se K-edge are consistent with those for the Fe K-edge, in which peak $D$ is due to 1$s$ → 4$p$ dipole transition and feature $E$ is the result of multiple scattering of the photoelectrons with the nearest neighbors. The increase of $D$ with $x$ is in agreement with the increase of $A$ and the stronger hybridization of Fe 3$d$-Se/S 4$p$ bands. The S substitution dependence of the Fe and Se K-edge XANES features resembles those observed in high pressure measurement and is opposite to effects induced by Te substitution.[48,51] Increased $d$–$p$ hybridization is expected to promote delocalization.[52]

EXAFS presents a window into local bond distances around absorbing atom and could be described in the single-scattering approximation as:[49]

$$\chi(k) = \sum_i \frac{N_i S_0^2}{k R_i^2} f_i(k_i R_i) e^{-\frac{2R_i}{\lambda}} e^{-2k^2 \sigma_i^2} \sin[2k R_i + \delta_i(k)] \quad (1)$$

where $N_i$ is the number of neighboring atoms at a $R_i$ distance from the photoabsorbing atom, $S_0^2$ is the passive electrons reduction factor, $f_i(k,R_i)$ is the backscattering amplitude, $λ$ is the photoelectron mean free path, $δ_i$ is the phase shift of the photoelectrons, $k$ is photon wavenumber and $σ^2$ is the correlated

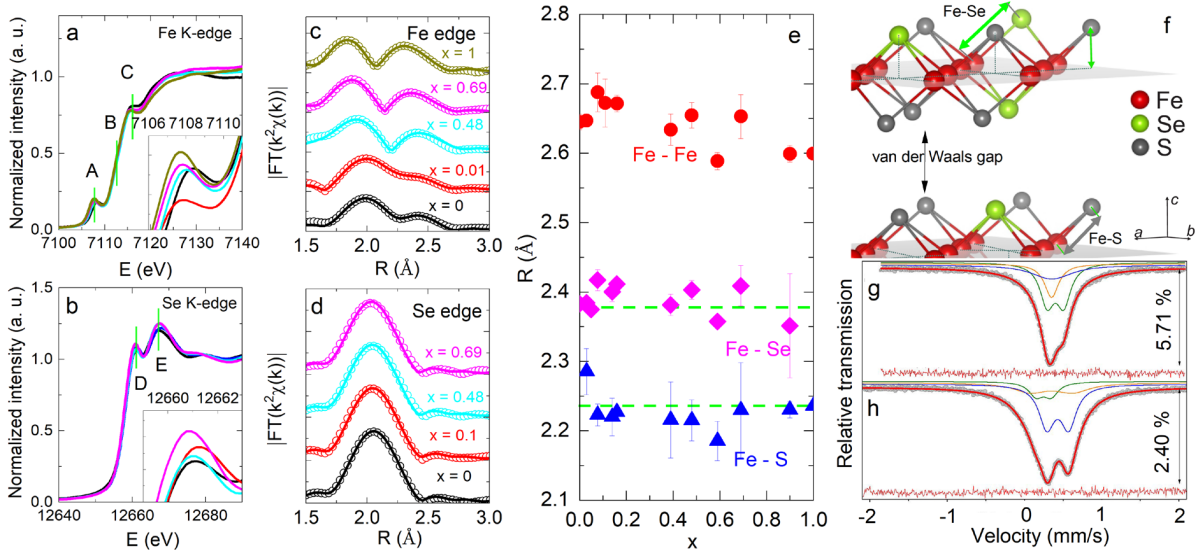

**Figure 4**. Local crystallography from XANES and EXAFS. (a,b) Normalized XANES spectra of $Fe_{1-y}Se_{1-x}S_x$ measured at Fe K-edge and Se K-edge. The features in the Fe K-edge XANES are denoted by *A*, *B*, *C*; Se K-edge are denoted by *D*, *E*. (c,d) Fourier transform (FT) magnitudes of the $k^2$-weighted EXAFS oscillations at Fe K-edge and Se K-edge. (e) Fe-Fe, FeSe, and FeS distances determined by EXAFS as a function of *x*. The Fe-Se and Fe-S bond distances show weak changes for all *x* and remain close to values observed in the end member compounds FeSe and FeS. (f) Crystal structure of $Fe_{1-y}Se_{1-x}S_x$ with depicted van der Waals gap, Fe planes and different Fe-Se and Fe-S bond distances. Mössbauer spectrum recorded in low velocity range at 295 K of the tetragonal $Fe_{0.96}Se_{0.31}S_{0.69}$ (g) and of the tetragonal $Fe_{0.9}Se_{0.82}S_{0.18}$ (h).

Debye-Waller factor measuring the mean square relative displacement of the photoabsorber-backscatter pairs. The first nearest neighbors of Fe atoms are four Se atoms located at 2.395 Å, and the next nearest neighbors are four Fe atoms sited at 2.668 Å.[39] The first nearest neighbors of Se atoms are four Fe atoms at 2.395 Å tetrahedral distances. Local structural information, such as the bond distance and Debye-Waller factor, were obtained by the best-fit model. The features above 3 Å are due to longer distances and multiple scattering effects. Since the first shell of the Se K-edge is well separated from the distant shells, we fit the Se K-edge EXAFS using a model with a single distance and take only the main peaks of the Fe-edge into consideration. Interestingly, the fitting can only be improved by using different values of Fe-Se and Fe-S bond lengths (Fig. 4e,f). Hence, Fourier transform magnitudes of the EXAFS oscillations weighted by $k^2$ and extracted from the Fe K and Se K-edges (Fig. 4c,d) indicate that Fe-Se and Fe-S bond lengths in all investigated materials, including in alloyed crystals, do not deviate much from Fe-Se and Fe-S bond lengths in pure FeSe and pure FeS. The Fe-Fe bond length shows a weak increase for $x \leq 0.10$, along with that of $T_c$ (Fig. 4e) and a decrease for further S substitution results as $x \rightarrow 1$. This suggests rather deformed $FeCh_4$ tetrahedra. Moreover, wider distributions of Fe-S and Fe-Se bond distances for $0.4 \leq x \leq 0.9$ inferred from error bars (Fig. 4e) imply higher disorder. Consequently, there is a local structure inhomogeneity in most $Fe(Se/S)_4$ tetrahedra the middle of the alloy series due to the comparable numbers of Fe-Se and Fe-S bonds in $FeCh_4$.

For insight into the local charge distribution around Fe atoms at a *2a* Wyckoff site of the *P4/nmm* space group we have used Mössbauer spectroscopy. First, we consider the coordination spheres of the iron probe. The four nearest neighbors (NN) chalcogen atoms make the surrounding $FeCh_4$ (Ch=Se,S) tetrahedron. The next-nearest-neighbors (NNN) are two metal shells with four iron atoms in each shell. Farther behind are two chalcogen shells which consist of four and eight chalcogen atoms, respectively. The nearer four are situated in the same atomic plane with the iron. The first half of farther eight atoms are above and the second half are below this plane stacked along the *c*-axis. The symmetry point group of the Fe atomic site in FeS and FeSe is $\bar{4}m2$. The non-ideal tetrahedral surrounding of Fe results in a non-spherical charge distribution around the probe and the emergence of an electric field gradient (EFG).[53,54] Substitution of different chalcogen atoms in an iron-chalcogenide material additionally breaks the Fe local symmetry, causing more pronounced EFG. Replacement of sulfur by selenium atoms defines a new Se-containing atomic plane which is different from iron and sulfur planes in FeS.[53] In pure materials, the bond distance $d_{Fe-Se}$ in tetragonal FeSe is ~10% longer than $d_{Fe-S}$ in tetragonal FeS.[54] The Se-plane in FeSe is farther away from the Fe-plane than the S-plane in FeS, expanding the crystallographic unit cell along the *c*-axis.

In alloys the number of occupied chalcogen sites in the Se or S planes depends on the Se/S atomic ratio. This leads to many nonequivalent Fe sites in contrast to the uniform chalcogen surrounding in pure FeS or FeSe. It is expected that Mössbauer spectroscopy will detect a distribution of quadrupole splittings (Δ) because of different EFGs.

The binomial distribution describes the probability for the appearance of coordination spheres of the Fe atom with different content of the S and Se ions:[55]

$$p(z_1, \ldots z_k, n_1, \ldots \ldots n_k, c) = \prod_{i=1}^{k} p_{z_i}(n_i, c) = \prod_{i=1}^{k} \binom{z_i}{n_i} c^{n_i}(1-c)^{z_i-n_i} \quad (2)$$

where $0 \leq n_i \leq z_i$ and p is probability mass function (PMF). PMF describes probability of having $n_i$ impurities on $z_i$ host sites of the *i*-th shell for impurity concentration c. Considering only the tetrahedron with four NN for c = 0.3, the equation gives approximately the ratio p($n_1$ = 0):p($n_1$ = 1):p($n_1$ = 2):p($n_1$ = 3):p($n_1$ = 4) = 24:41:26:8:1. Hence, the Δ-distribution by means of the doublets with Lorentz lines with area ratio follows the PMF ratio. The experimental spectra can only be fit with the three doublets fitting model where one of the doublets is collapsed into a single line in the spectrum of the tetragonal Fe$_{0.96}$Se$_{0.31}$S$_{0.69}$ (Table S2 in Supporting Information). This is to be expected since it originates from the tetrahedron without Se atoms where a very low Δ value was measured.[46,53] When S substitutes Se in FeSe$_4$ the subspectrum with the largest area is assigned to the FeSe$_4$ tetrahedron.[56,57] The other two doublets in both spectra arise from Se atom in FeS$_4$ tetrahedra, or vice versa, the existence of at least one S on the FeSe$_4$ tetrahedron corners.

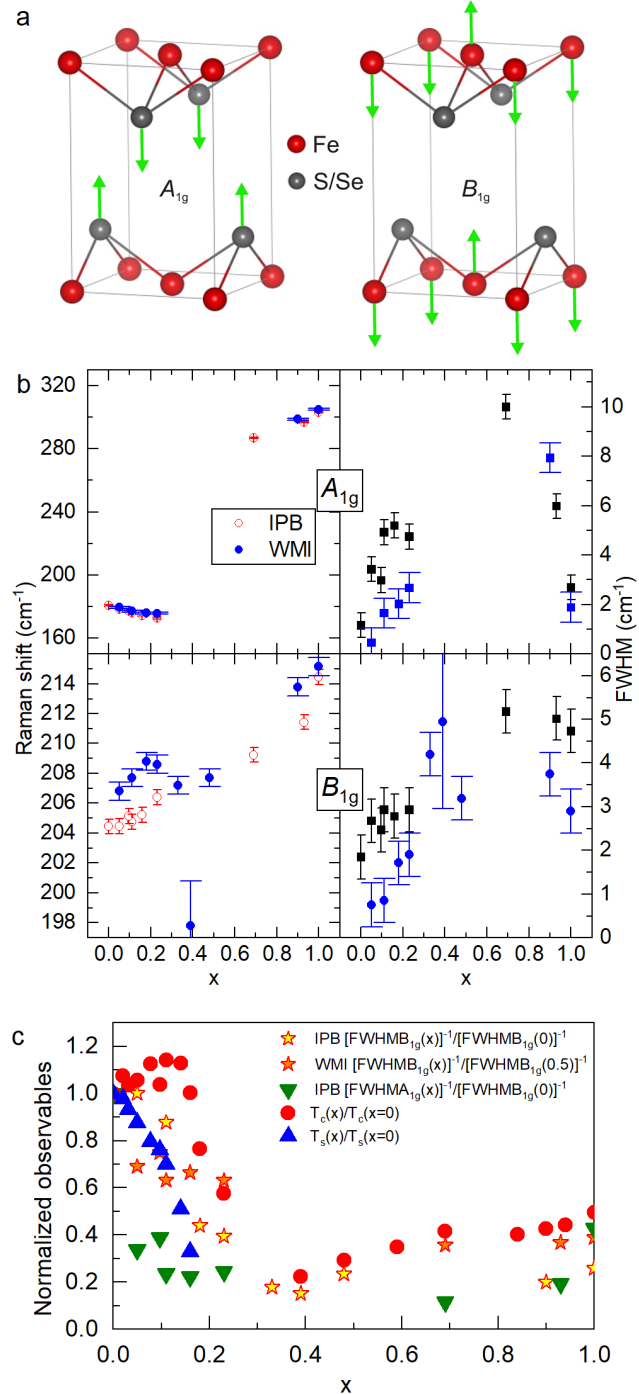

Wider lines and the large hyperfine parameter uncertainties are consequences of a great number of spatial combinations of substituted atoms. Bader analyses of the pure iron chalcogenides shows that there is a larger charge transfer from the Fe atom to the S atom than to the Se atom.[54] The charge transfer changes the chemical shift, i.e. $\delta$. Therefore, the value of $\delta$ is the result of a competition between electron occupation on 4$s$ and 3$d$-orbitals of the Fe.[58] According to the Bader analysis there is a bond critical point between the NN chalcogen ion and the NNN chalcogen from the adjacent layer in the fifth shell.[54] There are significant probabilities for the next PMF: p($n_1$ = 0, $n_s$ = 2,3), p($n_1$ = 1, $n_s$ = 1,2,3,4), and p($n_1$ = 2, $n_s$ = 1,2,3). All contribute to changes in the probe electron density, additionally smearing the quadrupole distribution with consequences on the subspectrum area ratio. Therefore, our data confirm the randomness of the spatial distribution of the Se and S atoms in alloys and therefore sulfur entails both pressure and disorder. Since there is no comparably dramatic increase in $T_c$ as in FeSe under pressure,[17] it is plausible to conclude that disorder counterbalances pressure for $x$ up to the NCP where the abrupt change of superconducting gap takes place.[19]

It is obvious that random sulfur substitution in FeSe and inhomogeneous Fe-Se/S bond distances in FeCh$_4$ tetrahedra induce changes of S and Se atomic heights from the average values inferred from Rietveld measurement (Fig. 2e). Therefore, we focus next on the atomic vibrations along crystallographic $c$-axis. Of interest are Raman shift and Lorentzian full width at half maximum (FWHM) of A$_{1g}$ and B$_{1g}$ Raman modes at 100 K (Fig. 5a). We observe a good agreement in trends of change of Raman shift and peak widths. In the absence of Fano-shape peak distortions, peak width gives a caliper of phonon vibrations-related crystallographic disorder. Larger FWHM is consistent with higher disorder. For better comparison with changes in superconducting $T_c$, in Fig. 5b we plot [FWHM($x$)]$^{-1}$, normalized to [FWHM (0)]$^{-1}$, i.e. to FeSe without S substitution, for B$_{1g}$ and A$_{1g}$ modes as S enters the lattice with increase in $x$. Smaller [FWHM($x$)]$^{-1}$/[FWHM(0)]$^{-1}$ point to increased disorder relative to pure FeSe.

The Raman active A$_{1g}$ and B$_{1g}$ phonon modes correspond to fully symmetric and the out-of-phase vibration of the chalcogen and Fe atoms along the $c$-axis, respectively.[59-61] The B$_{1g}$ mode has the same symmetry as the nematic fluctuations, both charge and spin, and is therefore expected to couple to deformations of this type.[60,62,63] Both lines broaden due to disorder as $x$ is increased. However, whereas the relative change in inverse half-width of the B$_{1g}$ phonon [FWHM B$_{1g}$ ($x$)]$^{-1}$ tracks relative changes in $T_c(x)$ and $T_s(x)$ (Fig. 5c), relative change in [FWHM A$_{1g}$ ($x$)]$^{-1}$ does not.

DISCUSSION

To start with, we note that the concept of anion height comes from the average crystal structure; it does not contain information from local crystallography. However, bond lengths regulate the Fe-Ch overlap and the FeCh$_4$ tetrahedron shape which, in turn, controls the crystal field levels and thus the orbital occupancies and relative mixing of Fe atom $d_{xz}$ and $d_{yz}$ orbitals.[11,12] Hence, disorder in Fe-Ch bond distances is likely to suppress $T_s$ which is rather sensitive on energy splitting between $d_{xz}$ and $d_{yz}$ atomic orbitals of Fe atom.[60,64] Moreover, it will also influence superconducting gap at the Γ point at the Brillouin zone via perturbation of $d_{xz}$ spectral weight at the Fermi level, and thus $T_c$.[65] Finally, high bond disorder in this context could be connected to strong charge-nematic or magnetic spin fluctuations that mediate superconducting pairing in the middle of the alloy series.[7,12,63,66] The increase in Fe vibrations disorder along the $c$-axis is coincident with the suppression in $T_c(x)$, as observed in Fig. 5c.

**Figure 5**. (a) Unit cell of FeSe/S with depicted A$_{1g}$ and B$_{1g}$ Raman modes. (b) Raman shift and Lorentzian full width at half maximum of the A$_{1g}$ and B$_{1g}$ modes. (c) Relative change of $T_c$, $T_s$ and widths of Raman active A$_{1g}$ and B$_{1g}$ modes with respect to FeSe; B$_{1g}$ mode measured at WMI is normalized to $x$ = 0.05.

A picture emerges where distinct Fe-S and Fe-Se chemical bonds randomly distributed in the lattice influence atomic vibration of Fe and Ch atoms in the $c$-axis direction. It would be of interest to determine the spatial extent of this inhomogeneity as well as local conducting properties by transmission electron and scanning tunneling microscopy. On the other hand, connection of Fe $B_{1g}$ mode disorder with the change of structural nematic and superconducting critical temperature with $x$ may affect high magnetic field properties such as upper critical field $H_{c2}$. This would be of interest for materials with higher $T_c$ such as FeSe crystals intercalated by molecular spacer layer or thin films.[27,28] We note that FeTe$_{1-x}$Se$_x$, amenable for wire fabrication and of interest for applications, also exhibits distinct Fe-Te and Fe-Se bond distances.[67,68] In fact, changes of Fe-Se and Fe-S bonds with $x$ (Fig. 4e) are smaller when compared to changes in Fe$_{1+y}$Te$_{1-x}$Se$_x$ alloys where Fe-Te and Fe-Se bond lengths exhibit only weak decrease from FeTe and FeSe.[69-71] The local structure inhomogeneity should have direct implications for the electronic states near the Fermi energy since they are connected with the degree of delocalization of the Fe $d$ orbitals.[12]

CONCLUSION

In summary, in Fe$_{1-y}$Se$_{1-x}$S$_x$ ($0 \leq x \leq 1, y \leq 1$) crystal alloys superconducting $T_c$'s do not scale with the anion height above the Fe plane in the presence (FeSe) or in the absence (FeS) of nematic order, in contrast to most known Fe-based superconductors. Investigation of the local crystallography from EXAFS and Mössbauer spectroscopy reveal that sulfur substitutes Se randomly in alloys whereas Fe-Se and Fe-S bonds remain the same as in pure FeSe and pure FeS, respectively. This causes disorder in atomic vibration orthogonal to Fe plane as seen in $A_{1g}$ and $B_{1g}$ Raman active modes of Se and Fe, respectively. Changes of $T_s(x)$ and $T_c(x)$ compared to $T_c$ and $T_s$ in FeSe are not connected with $A_{1g}$, but are connected with $B_{1g}$ mode disorder. Suppression of nematic transition associated with tetragonal-to-orthorhombic change of the crystallographic unit cell on cooling by S substitution on Se atomic positions follows the increased $B_{1g}$ disorder. In contrast, increased $B_{1g}$ mode disorder does not affect $T_c$ that rises with S substitution in the nematic region due to chemical pressure effect. For high bond disorder near the middle of the alloy series when $T_s$ is fully suppressed, relative changes in $T_c(x)$ with respect to FeSe follow relative disorder in $B_{1g}$ mode vibrations of Fe atoms. Since $B_{1g}$ mode couples to spin and charge nematic fluctuations, our results show that such fluctuations are tied to interlayer crystallography of FeSe/S and are of interest for design of novel intercalated and ultrathin FeSe materials.

ASSOCIATED CONTENT

**Supporting Information**. This material is available free of charge on the ACS publications website at DOI: http://pubs.acs.org.
Experimental details. (DOC)
Crystal Structure. (DOC)
Sample to sample reproducibility and Wilson ratio. (DOC)
First-principle calculation results. (DOC)

AUTHOR INFORMATION

Corresponding Authors

\* petrovic@bnl.gov
\* afwang@cqu.edu.cn

Present Addresses

&Present address: School of Physics, Chongqing University, Chongqing 400044, China.
#Present address: Materials Science Division, Argonne National Laboratory, Lemont, Illinois 60439, USA.
▽Present address: Los Alamos National Laboratory, MS K764, Los Alamos NM 87545, USA.
⊥Present address: ALBA Synchrotron Light Source, Cerdanyola del Valles, E-08290 Barcelona, Spain.

Notes
The authors declare no competing financial interest.

ACKNOWLEDGMENT

Work at Brookhaven National Laboratory was supported by US DOE, Office of Science, Office of Basic Energy Sciences (DOE BES), under Contract No. DE-SC0012704 (AW, YL, QD and CP). This research used the 8-ID (ISS) and 28-ID-1 (PDF) beamlines of the National Synchrotron Light Source II, a U.S. DOE Office of Science User Facility operated for the DOE Office of Science by Brookhaven National Laboratory under Contract No. DE-SC0012704. Raman spectroscopy measurements were funded by the Serbian Academy of Sciences and Arts contract F-134, by the Institute of Physics Belgrade through the grant by the Ministry of Education, Science and Technological Development of the Republic of Serbia and by the Science Fund of the Republic of Serbia, PROMIS, No. 6062656, StrainedFeSC. Mössbauer spectroscopy and corresponding calculations were funded by the Ministry of Education, Science and Technological Development of the Republic of Serbia

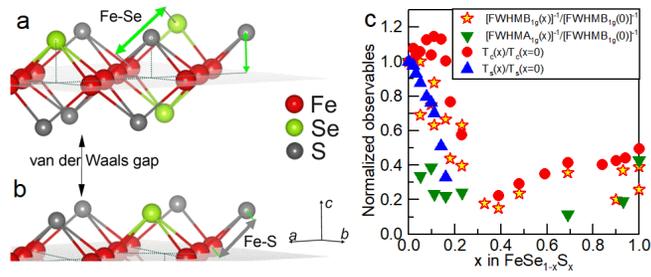

SYNOPSIS

The relation between critical temperature $T_c$ and crystal chemistry is reported for FeSe/S superconductor. In $Fe_{1-y}Se_{1-x}S_x$ ($0 \leq x \leq 1$, $y \leq 0.1$) not only $T_c(x)$ but also tetragonal-to-orthorombic transition $T_s(x)$ are correlated with disorder in Fe vibrations along the crystallographic $c$-axis, caused by the different Se/S anion height above the Fe plane. The importance of local crystal structure opens new window for predictive superconductor design.